# Considerations for Single-Arm Trials to Support Accelerated Approval of Oncology Drugs

Feinan Lu[1], Tao Wang[1], Ying Lu[2], Jie Chen[1]*

[1]ECR Global, Shanghai, China

[2]Department of Biomedical Science, Stanford University School of Medicine, Stanford, California, USA

## Contents

*Correspondence to: Dr. Jie Chen, ECR Global, Shanghai, China (jie.chen8@ecr-global.com).

# Considerations for Single-Arm Trials to Support Accelerated Approval of Oncology Drugs

**Abstract**

In the last two decades, single-arm trials (SATs) have been effectively used to study anticancer therapies in well-defined patient populations using durable response rates as an objective and interpretable clinical endpoints. With a growing trend of regulatory accelerated approval (AA) requiring randomized controlled trials (RCTs), some confusions have arisen about the roles of SATs in AA. This review is intended to elucidate necessary and desirable conditions under which an SAT may be considered appropriate for AA. Specifically, the paper describes (1) two necessary conditions for designing an SAT, (2) eight desirable conditions that help either optimize the study design and doses or interpret the study results, and (3) three additional considerations for construction of estimands, adaptive designs, and timely communication with the regulatory agency. Three examples are presented to demonstrate how SATs can or cannot provide sufficient evidence to support regulatory decision. Conditions and considerations presented in this review may serve as a set of references for sponsors considering SATs to support regulatory approval of anticancer drugs.

**Keywords**: Rare cancers, natural history study, mechanism of action, dose optimization, substantial evidence

## 1 Introduction

The United States Food and Drug Administration (FDA) established in 1992 accelerated approval (AA) that later led to the expedited programs for serious conditions in several disease areas (FDA, 2014a). In the guidance for expedited programs, the FDA states that "...single-arm trials may be an important option in rare diseases with well-understood pathophysiology and a well-defined disease course" (FDA, 2014a). One of the most important applications of single-arm trials (SATs) is the development of novel anticancer therapeutics for AA or conditional marketing authorizations (CMAs). Studies show that regulatory approvals based on SATs account for (1) 49% of FDA's AAs between 1992 and 2020, most of which (47% among all AAs) for oncology indications (Ribeiro et al., 2023), (2) 34% of CMAs by the European Medicines Agency (EMA) between 2006 and 2016 (Marcal, 2023) (20% for anticancer therapies between 2014 and 2016 (Naci et al., 2019)), (3) 42% of oncology drug approvals by the China National Medical Products Administration (NMPA) between 2018 and 2022 (Zhang et al., 2023), and (4) 21% of oncology drug approvals in Japan between 2006 and 2019 (Hatogai et al., 2021). Following AAs or CMAs, confirmatory trials are usually required to verify the intended clinical benefits, and failure to do so may cause withdrawal of the product. Studies indicate that there has been an increasing number of withdrawals of oncology products that were approved based on SATs, especially in the past decade or so Ribeiro et al. (2023); Cramer et al. (2023); Koole et al. (2024); Mellgard et al. (2024). For example, Agrawal et al. (Agrawal et al., 2023) report that among 116 FDA approved oncology indications based on SATs between 2002 and 2021, 61 (52%) are pending verification of clinical benefits and 10 (9%) have been withdrawn. An analysis of FDA's AA database (`https://www.fda.gov/drugs/nda-and-bla-approvals/accelerated-approval-program`) indicates a 13% withdrawal rate for FDA's AAs based on SATs between January 2017 and April 2023.

Recently the FDA issued a draft guidance on clinical trial considerations to support AA of oncology products (FDA, 2023a), in which the agency states that "… a randomized controlled trial is the preferred approach to support an application for accelerated approval… there can be circumstances wherein a single-arm trial is appropriate in the development of a drug for accelerated approval, for example when there are significant concerns about the feasibility of a randomized controlled trial." The EMA issued in 2023 a draft reflection paper on establishing efficacy based on SATs (EMA, 2023) which describes primarily the following five aspects: endpoints, target population, external information, statistical principles, and potential biases and mitigation. In addition, the NMPA released a guidance on marketing authorization for anticancer drugs based on SATs (NMPA, 2023) that focuses on the applicability of SATs—unmet medical needs, mechanism of action, external control, significant anticancer activities, manageable safety concerns, and rare cancers.

Given the above background and considering the limitations (Grayling and Mander, 2016; Shah et al., 2021; Moon, 2022; Goldberg, 2021; Merino et al., 2023) and clear clinical values EMA (2023); NMPA (2023); Tenhunen et al. (2020); Agrawal et al. (2021); Mulder et al. (2023) of SATs, we intend in this paper to provide a comprehensive review from a statistical perspective on the necessary and desirable conditions under which an SAT may be appropriate to support an AA or CMA.

The paper is organized as follows. Section 2 describes necessary considerations for designing an SAT and Section 3 presents desirable conditions for using an SAT to support regulatory decision. Additional considerations for construction of estimands, adaptive designs, and communication with regulatory agencies are provided in Section 4. Three examples are presented in Section 5 and some concluding remarks are given in Section 6

## 2  Necessary Conditions

The FDA guidance on accelerated approval of oncology products outlines considerations in the design, conduct, and analyses of single-arm trials (FDA, 2023a). This section further elucidates the necessary conditions (including those in (FDA, 2023a; EMA, 2023; NMPA, 2023)) and the following section presents desirable conditions, based on which an SAT can be considered to support AA.

### 2.1  Life-threatening or serious conditions with no efficacious treatments

The majority of cancer types are life-threatening, especially those in terminal stages. Cancer patients who are relapsed or refractory after multiple prior-line therapies are generally in unmet medical needs with few treatment options and poor clinical outcomes (Lu et al., 2017). For example: (1) patients with relapsed or refractory diffuse large B-cell lymphoma who failed at least two multi-agent systemic anticancer treatment regimens generally have poor prognosis with limited treatments available (Caimi et al., 2021), (2) the majority of patients with extensive-stage small-cell lung cancer (SCLC) often have disease progression within 6 months even after front-line treatments (Rosner and Levy, 2023), and (3) patients with locally advanced or metastatic urothelial cancer have limited treatment options because many of them are ineligible for cisplatin-based chemotherapy (a standard of care or SOC) due to substantial toxicity (Hoimes et al., 2023).  All of these scenarios share the same or similar features of unmet medical needs—The cancers are life-threatening or serious conditions with limited or no treatment options, often resulting in unsatisfactory survival or poor quality of life for cancer patients. Therefore, it would be unethical to assign the patients to the control group in an RCT where there is no efficacious control treatment available.

## 2.2 Rare cancers

Rare cancers present unique challenges in drug development, which may include (1) difficulty to enroll insufficient number of patients in clinical trials due to small patient population size, (2) limited understanding of rare cancer pathophysiology and natural history, and (3) difficulty in designing and conducting RCTs due to lack of efficacious therapies available as a control. These challenges, together with others such as severity and early onset (often at childhood), are barriers to conduct an RCT; instead, an SAT with either an implicit or explicit external control, at the population level (e.g., a fixed threshold value for a response rate) or at individual patient level (e.g., propensity score matching), may be more appropriate to investigate the anticancer effect of a new drug (Hashmi et al., 2021; Oda and Narukawa, 2022).

# 3 Desirable Conditions

## 3.1 Well-understood natural history of the rare cancer

Understanding the natural history of a rare cancer is important to help (1) identify the right target population through defining appropriate eligibility criteria (including genotypes and phenotypes), (2) detect changes in the pattern of disease course and assess clinical outcomes, (3) develop biomarkers for diagnosis and prognosis of the disease and for choosing appropriate endpoints (FDA, 2019). For example, knowing the natural history of a rare cancer can help differentiate whether the disease progression or improvement of conditions is due to the treatment under investigation or the natural course of the cancer. For many cancers, it is known that major tumor regression may not occur without effective treatment; therefore, objective response rates are often used to reflect the direct drug effect as substantial evidence supporting AA, which can be very valuable for trials in refractory and/or relapsed cancers (Delgado and Guddati, 2021). Note that rare cancers often comprise a multitude of cancer subtypes affecting heterogeneous patient population, leading to even more difficult-to-fully-understand the etiology and natural history of many rare cancers (FDA, 2024). Therefore, it is important to conduct a natural history study before planning an SAT to characterize demographic, genetic/genomic, environmental, and other factors (e.g., with treatment of standard of care) that may correlate with the development and outcomes of the cancer.

## 3.2 Well-understood mechanism of action of the drug

Most modern anticancer treatments are targeted therapies that target proteins that control the growth, division, and/or spread of cancer cells. For example, a targeted therapy may (1) help the immune system destroy cancer cells, (2) stop cancer cells from growing by interrupting signals causing them to grow and divide, (3) stop signals that help form blood vessels, (4) deliver cell-killing substance to cancer cells, and (5) cause cancer cell death (apoptosis) (NCI, 2022). Understanding the mechanism of action (MOA) of a drug is essential for identifying surrogate markers (endpoints) of treatment effects, determining adequacy of dosage, selecting cancer patient (sub)population based on existence (or absence) of the target/receptor, and/or suggesting strategies for combination therapies. For example, (1) an anti-PD-1 (programmed cell death 1) ligand (PD-L1) drug blocking immune checkpoint pathway is effective against several cancer types, which helps identify subsets of patients who are likely responsive to an anti-PDL1 therapy (Topalian et al., 2016), (2) an anti-CD19 combined with anti-CD20 CAR-modified T cells for B-cell is effective in treating patients with hematological malignancies with both CD19 and CD20 antigens expressed on their B cells (Sang et al., 2019), and (3) an anti-drug conjugate, typically composed of monoclonal antibody, reaches the therapeutic target and then releases the cytotoxic payloads in the vicinity of the targets (Fu et al., 2022).

In recent years, several tumor-agnostic indications have received AA (Westphalen et al., 2024). For example, the FDA granted in May 2017 AA to pembrolizumab in adults and children affected by unresectable or metastatic solid tumors with deficient mismatch repair (dMMR) and/or high microsatellite instability (MSI-H), pretreated and without any valid alternative treatment option, where both dMMR and MSI-H are predictive biomarkers of response to PD1 blockade. Recently the European Society for Medical Oncology (ESMO) issued a new framework to assess tumor-agnostic potential of anticancer therapies, in which ESMO proposes the following pragmatic conceptual basis with three categories for assessing to the therapeutic effect of investigational therapies: (1) *tumor-agnostic* when targeting a driver molecular aberration predominantly defines the therapeutic effect, irrespective of tumor-specific biology (TSB), (2) *tumor-modulated* when the therapeutic effect on a targeted driver molecular aberration is modulated by the TSB (e.g., PARP inhibitors in tumors harboring BRCA1/2 mutation/homologous recombination deficiency), and (3) *tumor-restricted* when the therapeutic effect on a targeted driver molecular aberration is only present in a TSB context (e.g. PI3K inhibitors in PIK3CA-mutated breast cancer) (Westphalen et al., 2024). This classification of anticancer therapies is useful in elucidating the therapeutic effect and associated MOA of molecularly guided treatment options (MGTOs).

### 3.3 Adequate dose optimization

Traditional first-in-human dose-escalation studies determine the maximum tolerable dose (MTD), or a dose close to the MTD, which is often recommended for subsequent clinical studies without further dose optimization. This approach may well be suited to cytotoxic agents such as chemotherapies, but may not be appropriate for target therapies (e.g., kinase inhibitors, monoclonal antibodies, and anti-drug conjugates) that interact with a molecular pathway and that demonstrate different dose-response relationship. The MTD approach is based only on a short period of observations with a limited number of subjects and ignores target interactions and off-target toxicities (Moon, 2022; FDA, 2023b). With target therapies, increasing doses beyond certain level may not improve anticancer activities, and serious and intolerable adverse effects may occur after a multi-cycle, persistent treatment, leading to dose interruption and reduced compliance (Shah et al., 2021).

The goal of dose optimization is to identify a dose or a dose range that produces the maximum possible efficacy while maintaining acceptable toxicity. This is usually achieved by randomized dose optimization trials (DOTs) that focus on the relationship of drug exposure with antitumor activities including both efficacy and toxicity. However, the traditional stagewise approach for dose finding and optimization may potentially exclude the true optimal dose(s) in stage 1 and usually requires relatively large sample sizes (Yuan et al., 2024a), which may not be feasible for rare cancers. To address these limitations, some alternative strategies such as the followings may be considered in dose optimization for drugs treating rare cancers:

- Efficacy-integrated approach: The phase I dose finding study can make the use of both toxicity and efficacy to explore the exposure-response relationship and to guide dose escalation decisions through benefit-risk trade-off (Liu and Johnson, 2016), in which dose optimization may be achieved through the use of toxicity endpoints and short-term efficacy endpoints, e.g., pharmacometric endpoints or target receptor occupancy, to guide nearly real-time dose decisions and then the use of long-term clinical efficacy endpoints at the end of the trial for identifying optimal biological dose(s) (Yuan et al., 2024a,b; Agrawal et al., 2016; Zhou and Ji, 2024).

- Seamless phase I-II designs: Examples of this type of designs may include BARD (backfill and adaptive randomization for dose optimization) that may reduce the sample size (Zhao et al., 2024), DROID (a dose-ranging approach to optimizing dose) that bridges the dose-ranging framework with oncology dose-finding designs to estimate the optimal dose(s) (Guo and Yuan, 2023), PEDOOP (pharmacometrics-enabled dose optimization) that incorporates patient-level pharmacokinetics and latent pharmacodynamics information for dose optimization (Yuan et al., 2024b), and intra-patient dose escalation to reduce the number of patients being exposed (Wirth et al., 2020). These designs have the potential to reduce the sample size and improve the accuracy of identifying the true optimal dose(s).

- Basket-nested designs: For cancers with the same or similar target, one may consider expansions of multiple doses in multiple tumor types (basket-nested design) after dose escalation and using Bayesian methods to borrow evidence from other tumors for dose optimization (Zhou and Ji, 2024; Jiang et al., 2024; Lyu et al., 2023; Wang et al., 2024).

In addition to the above strategies and the general recommendations in regulatory guidance (NMPA, 2023; FDA, 2023b), some other considerations in such DOTs may include heterogeneity of patient population (tumor type, disease stage, comorbidities, etc.), the selection of the most appropriate endpoints that can accurately capture antitumor activities (e.g., anticancer response or progress-free survival), therapeutic properties of the drug (e.g., small or large molecules, agonist or antagonist), and whether there are approved indications for the drugs (Dejardin et al., 2024; Korn et al., 2023; Papachristos et al., 2023).

## 3.4 Substantial treatment effects

SATs often use surrogate endpoints (SEs), such as objective response (OR), complete response (CR), and duration of response (DOR), to gauge anticancer activities of drugs. For example, most oncology SATs use OR rate (ORR) and/or CR rate (CRR) as primary measures and DOR as a key secondary measure of anticancer effects. For SATs seeking AA, the trial should generate an ORR (and/or CRR) that is *clinically meaningful and statistically significantly higher* than the rate that the same patient population would experience without taking the drug or with some SOC (FDA, 2023a). In addition, the DOR among responders should be long enough to ensure that the observed responses are due to treatment, not to the natural history of the cancer or bias of patient selection. Studies have showed that modest improvement on SEs may not last long enough and/or translate into clinically meaningful survival benefit of cancer patients (Goldberg, 2021; Lemery and Pazdur, 2022; Ribeiro et al., 2022; Mullard, 2023). Of note, substantial treatment effects can be concluded by comparing the effects from the SAT with those from a pre-specified external control (e.g., a historical or contemporaneous control (Mishra-Kalyani et al., 2022)) at the individual patient level or with a fixed value for an ORR and/or DOR).

There are other intermediate efficacy endpoints such as disease-free survival, time-to-progression, probability of response maintenance, progression-free survival (PFS), and relapse-free survival for solid tumors. These endpoints are usually measured as secondary endpoints, each of which reflects some specific aspects of treatment effect and should also be considered together with the primary and/or key secondary endpoints in an SAT (FDA, 2018). Note that complete remission or complete remission with (partial) hematologic recovery (of peripheral blood counts) (CRh), major molecular response, cryogenic response, and/or minimal residual disease are often used for regulatory decisions in developing drugs treating hematologic malignancies (FDA, 2020).

In general, assessment of substantial evidence should take into account the followings: (1) the magnitude of estimated effects in terms of selected (primary and key secondary) endpoints, (2) the target patient population (e.g., patients with terminal diseases or prior two or more lines of therapies), (3) external information about natural history, external controls, SOC, and unmet medical needs, (4) statistical considerations with respect to sample size, hypothesis testing, multiplicity, missing data, and sensitivity analysis, and (5) rigorous supplementary and sensitivity analyses to fully explore all possible biases (e.g., patient selection bias, endpoint assessment bias, attrition bias, immortal time bias) and their impact on the study conclusion. Whether an SAT can generate substantial evidence of treatment effects depends on the clinical context and should be discussed with relevant regulatory agencies prior to initiation of the SAT (FDA, 2023a; EMA, 2023; NMPA, 2023).

## 3.5 Translation of SEs into clinical benefits

Clinical benefits are commonly measured by prolongation of survival (e.g., PFS or overall survival (OS)) or improvement in quality of life of cancer patients. The relationship between SEs and survival has not been formally established in many cancer types treated with anticancer agents and may depend on many factors such as the stage of the cancer, number of prior-line therapies, magnitude of treatment effect as measured by SEs, and safety profile of the drug and its MOA. Merino et al. (Merino et al., 2023) point out that ORR may not translate into overall survival benefit because (1) a modest magnitude of ORR may be transient or due to the natural history of the disease or patient selection bias and/or (2) a large ORR often accompany with a higher (suboptimal) dose that aims at producing a higher ORR but may be associated with intolerable toxicity (that causes early withdrawal with shorter follow-up time or even significant treatment-related deaths) (Bou Zeid and Yazbeck, 2023). On the other hand, the relationship between early tumor-based SEs and OS can be bidirectional, some trials have showed OS benefit without substantial improvement in ORR (Prasad et al., 2015; Burtness et al., 2019), which may be caused by the unique MOA for target therapies (Merino et al., 2023).

While the correlation between SEs and clinical benefits depends on the clinical context, the drug under investigation, the endpoint selected, and other design aspects (e.g., patient selection), some additional considerations in assessing translation of SEs into clinical benefits may include: (1) the magnitude of treatment effects as measured by the surrogates–the larger the observed effect, the higher the likelihood (in general) for the SEs to be translated into OS, (2) an established strong correlation of CR with OS in some cases (Huang et al., 2024), (3) consistency in the components of composite SEs (e.g., ORR and DOR), (4) understanding relationship among the SEs, biological plausibility, disease progression, and OS, and (5) consultation with regulatory agencies to ensure alignment on the use of SEs and their relevance to clinical benefit.

## 3.6 Favorable benefit-risk profile

The benefit-risk assessment (BRA) is quite complex in clinical trials, requiring considerations on multiple aspects, e.g., analysis condition, alternate treatment options, benefits, risks, and risk management (FDA, 2023c). It is even more challenging in SATs because of no comparison arm and difficulties to assess clinical benefits (due to shorter follow-up time) and to balance the observed antitumor effect with toxicity. However, the following considerations may be helpful in the BRA of an SAT:

- *Benefits*. SEs are commonly used in BRA of SATs Tenhunen et al. (2020); Mulder et al. (2023), in which some tools such as the magnitude of clinical benefit scale (ESMO, 2023) can be helpful. Although the strength of correlation among SEs, intermediate endpoints, and clinical endpoints depends on the clinical context Daniele et al. (2023), in general, (1) a durable response rate is usually associated with a better likelihood of having clinical benefits, (2) CR may be a better option than ORR in predicting OS, and (3) intermediate endpoints (e.g., PFS) may be better predictors of OS than ORR.

- *Risks*. Assessment of risk/harms in SATs should focus on (1) observed AEs and their clinical importance, (e.g., severity, frequency of occurrence, tolerability), (2) level of certainty for causality, and (3) manageability of the risks (FDA, 2023c; Califf, 2017). Note that risk assessment is particularly challenging in SATs because (1) symptoms of the disease are often prominent and in many cases indistinguishable from drug-induced AEs and (2) true incidences of drug-induced AEs are inestimable due to the absence of a control group.

The BRA can be performed quantitatively using a structured and well-defined process (FDA, 2023c; IMI-PROTECT, 2023). Note that the BRA for SATs should take into account the clinical context (e.g., the consequence if patients do not receive the treatment under investigation) and often the benefit-risk profiles of external controls (if any).

## 3.7 Totality of evidence

Regulatory approvals of medical products are based on the totality of evidence on the product effectiveness. In addition to the evidence discussed in Sections 3.4–3.6, the following aspects may be considered to supplement the totality of evidence on product effectiveness (FDA, 2023d):

- *Evidence from preclinical studies.* Evidence of efficacy and toxicity from *in vitro* and *in vivo* studies using the same endpoints that may translate to a similar clinical outcome (ICH, 2009).

- *Evidence from pharmacokinetic and pharmacodynamic (PK/PD) studies.* PK/PD data are required in almost all regulatory submissions to help understand the dose-response relationship, MOA of the drug, and disease pathophysiology (Araujo et al., 2023).

- *Evidence from other indications.* For indication expansion of approved drugs, it is important to integrate all data at all stages from other (approved and unapproved) indications, especially those with related indications or MOA, which may provide further evidence on the plausibility of treatment benefits and safety information (Gao et al., 2023).

- *Evidence from studies of pharmacologically similar products.* Evidence from approved products in the same pharmacological class may include the MOA, treatment effects on the same endpoints, and consistency of treatment effects across class members (Martini et al., 2017).

- *Real-world data (RWD) and real-world evidence (RWE).* External data, either at indication/study level or at individual patient level, are often used to determine the effectiveness of the drug, e.g., registry studies for natural history of a rare tumor and response rate for a terminal cancer treated with SoC. In many cases, RWD are used to select patients for external comparison group, for which there is extensive literature on consideration in the design, conduct, analysis, and result interpretation of externally controlled trials (Chen et al., 2023; FDA, 2023e).

- *Evidence from expanded access.* In some cases, patients with serious or immediately life-threatening rare cancers that lack effective treatments may be offered the investigational product via an expanded access program (EAP). Evidence derived from such an EAP can provide effectiveness and safety information in support of regulatory decision (Kempf et al., 2018; Feinberg et al., 2020).

## 3.8 Planning/initiation of confirmatory trials

For anticancer drugs granted AA, post-approval confirmatory trials are required to verify and describe anticipated clinical benefit (FDA, 2023a). Therefore, timely planning and/or initiation of a confirmatory trial is essential to support subsequent full approval within a reasonable time. Such a confirmatory trial (often an RCT) is intended to address the uncertainty about the relationship between SEs and clinical benefits (often measured by survival). Details about the design and initiation of the confirmatory RCT should be discussed as early as possible with relevant regulatory agencies (FDA, 2023a; EMA, 2023; NMPA, 2023).

# 4 Other Considerations

Besides the conditions described in the previous two sections, additional considerations may help well shape up an SAT from the trial design and communication perspectives.

## 4.1 Well-defined estimands

Upon meeting the criteria described in Section 2 and 3.1–3.3, then an SAT can be designed to demonstrate treatment effects of the drug by clearly defining the study objective(s) and corresponding estimands, with the latter often being described through precise definitions of estimand attributes (EMA, 2023; ICH, 2021). Using ORR as an example, the five attributes can be described as follows:

- Population: Patients with the cancer type of interest, possibly biomarker-defined

- Treatment: The new drug under investigation (and/or rescue therapies or SOC, if any)

- Endpoints: Responses (complete or partial response) by independent review committee (IRC), DOR

- Intercurrent events (ICEs): Discontinuation of the drug under investigation due to (1) intolerability, (2) start of another anticancer therapy (e.g., starts the next-line therapy because of disease progression (DP) by investigator's assessment while no DP by IRC), (3) unknown status (e.g., loss to follow-up), (4) terminal events (e.g., death)

- Population-level summary: Proportion of responses

Most of the ICE handling strategies can be used in SATs (Englert et al., 2023). For those seeking AA, the while-on-treatment strategy is often relevant as it concerns responses to the treatment before the occurrence of an ICE. However, the hypothetical strategy may also be applied if, e.g., a next-line therapy is used in patients who discontinue the originally assigned therapy for reasons other than toxicity and disease progression. Of note, for SATs with external controls, special attention should be given to precise definitions of target population (to minimize patient selection bias), treatment strategies (including rescue therapies), and different patterns of ICE occurrence between patients in the SAT and those in the external control (Chen et al., 2023, 2024). In general, the average treatment effect among the treated (ATT) estimand in an SAT is of interest to regulators.

## 4.2 Adaptive designs

Alternative to the follow-up confirmatory RCT discussed in 3.8, one may consider some innovative study designs that take into account both tumor responses and patient survival in different stages of a single study, e.g., a seamless adaptive design that models the response-survival relationship using pre-specified statistical methods (Lai et al., 2012) and a two-stage transition design, in which the first stage is an SAT and the second stage is an RCT (Shi et al., 2020). Adaptive designs can also be considered for confirmatory basket trials for multiple cancer types based on molecular alterations or biomarkers. Adaptive basket trials, often comprising multiple single-arms with each corresponding to a single tumor (sub)type, are particularly useful for rare cancers and have led regulatory approval for uncommon molecular alterations; see Beckman et al. (Beckman et al., 2016), Woodcock and LaVange (Woodcock and LaVange, 2017), Yu et al. (Yu et al., 2023), and Subbiah et al. (Subbiah et al., 2024) for examples of regulatory approvals based on adaptive basket trials.

### 4.3 Communication with regulatory agencies

It is highly recommended that sponsors communicate with relevant regulatory agencies on the suitability of an SAT to support AA before initiation of the study (NMPA, 2020; Zou et al., 2023). In the communication, the sponsors may consider at least the following aspects: (1) evidence from prior studies (including trial design, study population, treatment regimen, sample size, and endpoints); (2) safety and tolerability information (including dose limiting toxicity), AEs (especially SAEs), major organ toxicity, and dose-exposure-response characteristics; (3) clinical pharmacological data (including single- and multi-doses PK/PD data); (4) efficacy evidence (including target indication), all efficacy-related endpoints under the recommended dose for the target indication; (5) biomarker validation results if biomarker is used to define the target population; (6) background information about the target indication (including comprehensive review, incidence/prevalence, and current treatment options and associated effectiveness); (7) description of potential unmet medical needs of the investigational product for the target indication; (8) rationale for an SAT design (including complete protocol with eligibility criteria, primary and secondary efficacy endpoints, treatment regimen, sample size, statistical hypothesis) and planning for the confirmatory RCT; and (9) relevant information about the use of independent review committee and its charter.

## 5 Examples

This section presents three cases with each representing (1) from AA to full approval, (2) AA failure, and (3) first AA and then withdrawal indications based on later confirmatory trials.

### 5.1 Blinatumomab for relapsed/refractory B-cell acute lymphoblastic leukemia

Prognosis for patients with Philadelphia chromosome–negative (Ph-neg) relapsed or refractory B-precursor acute lymphoblastic leukaemia (R/R ALL) who have failed multiple lines of therapies is usually unfavorable with median overall survival (mOS) of 3–6 months (Gökbuget et al., 2012). The B-lineage surface antigen CD19 is homogeneously and stably expressed in over 95% of B-precursor ALL blasts, making it a likely target for immunotherapy (Raponi et al., 2011). Blinatumomab is a bispecific CD19-directed CD3 T-cell engager antibody that simultaneously binds CD3-positive cytotoxic T cells and CD19-positive B cells, leading to T-cell-mediated serial killing of malignant B cells (Hoffmann et al., 2005).

A phase II SAT (NCT01466179) enrolled 189 patients with Ph-neg R/R ALL. The primary efficacy endpoint was complete remission (CR) + CRh by two cycles of treatment. The study was designed to enroll at least 140 eligible patients to demonstrate that the rate of CR+CRh > 30% with 96% power at a one-sided significant level 0.025 if the true rate of CR+CRh is 45%. Blinatumomab was given to eligible patients by continuous intravenous infusion over 4 weeks of a 6-week cycle (9 ug/day for the first week and 28 ug/day thereafter for up to five cycles). The starting dose was based on the safety results of two prior studies and the step-dose regimen was found to be tolerable and active. At the time of primary efficacy analysis, all enrolled patients completed at least two cycles or discontinued early (Przepiorka et al., 2015; Topp et al., 2015). The analysis of primary endpoint showed that, using the treatment policy (intention-to-treat) strategy, the rate of CR + CRh was 43% with the lower bound of the 95% confidence interval [CI] being 35%. A weighted analysis of 694 historical control patients from 13 study groups and clinical centers showed an average CR of 24% with 95% CI being 20%–27% (FDA, 2014b), confirming the target lower limit of 30% CR + CRh for the accrued population.

The FDA granted in July 2017 a full approval of blinatumomab based a randomized, active controlled, open-label, phase III study investigating the efficacy of blinatumomab versus SOC chemotherapy in 405 adult patients with Ph-neg R/R B-cell precursor ALL. Results of the study showed a statistically significant improvement in OS for patients treated with blinatumomab (OS 7.7 months) compared to those treated with SOC (OS 4.0 months) (hazard ratio 0.71; 95% CI: 0.55, 0.93, *p*-value = 0.012). The approval also included results from a phase II study supporting the treatment of patients with Ph-positive R/R B-cell precursor ALL (FDA, 2017). See https://www.drugs.com/history/blincyto.html for the approval history of blinatumomab.

## 5.2 Retifanlimab for locally advanced or metastatic squamous carcinoma of the anal canal

Squamous cell carcinoma of the anal canal (SCCAC) is a rare cancer with an annual incidence rate ranging from 0.5 to 2.0 per 100,000 population worldwide (Morton et al., 2018). For relapsed and/or metastatic SCAC, there is no approved systemic treatment and chemotherapy is the standard of care with an average 5-year survival rate of only 30% (Rao et al., 2021), suggesting an unmet medical need.

Studies show that PD-1 and/or PD-L1 are expressed in over 50% SCCAC patients (Sharma et al., 2024). Therefore, immunotherapy with anti-PD-1 and anti-PD-L1 antibodies could be a potential effective therapy for SCCAC. Retifanlimab is a humanized, hinge-stabilized, immunoglobulin G4 kappa (IgG4k) monoclonal antibody that binds to and inhibits PD-1 and its downstream signal pathway, restoring immune function through the activation of T-cell and cell-mediated immune responses against tumor cells. Based on this MOA, a phase II SAT (POD1UM-202, NCT03597295) was designed to investigate the anticancer activities of retifanlimab among patients with locally advanced or metastatic SCCAC who have progressed after chemotherapy. Eligible patients received a 500-mg dose of retifanlimab every 4 weeks as an intravenous infusion (day 1 of each 28-day cycle) for up to 26 cycles. The primary efficacy endpoint was objective responses (either CR or PR) as evaluated by independent central review (ICR) and the secondary efficacy endpoints were DOR, disease control rate, PFS and OS. A sample size of 81 patients was planned to provide a 95% lower confidence limit of 13% with 80% probability if the true ORR was 24% (Rao et al., 2022).

The study enrolled 94 patients. At the data cutoff with a median follow-up of 7.1 months (range 0.9 – 19.4 months), 13 (1 CR + 12 PRs) patients had responses, 76 had discontinued treatment due to disease progression ($n$ = 58), AEs ($n$ = 6), death ($n$ = 6), physician decision ($n$ = 2), and lost to follow-up and withdrawal by patient (each $n$ = 1) and 18 patients were continuing the study treatment. The estimate of the primary efficacy endpoint using treatment policy strategy was 13.8% (95% CI: 7.6% – 22.5%) based on confirmed tumor responses by ICR. The estimated median DOR in responders was 9.5 months (95% CI: 5.6 months to inestimable). After reviewing the data, the FDA summarized the results in a pre-BLA (Biological License Application) meeting as follows: (1) ORR was modest and less than the pre-defined target of 25%, (2) DOR data was limited as only 7 of the 13 responders had DOR > 6 months, (3) if the BLA is submitted with results only from POD1UM-202, the agency might elect to discuss at an ODAC (Oncologic Drugs Advisory Committee) meeting, (4) the BLA would be a stronger application if it were supported by randomized controlled trials (POD1UM-303/InterAACT 2, NCT04472429, expected to be reported in 2025). On June 24, 2021, the FDA ODAC recommended with a 13-4 votes against AA of retifanlimab. Although the sponsor inferred the delay in disease progression, the agency insisted that SATs cannot produce such data because of lack of control group (Goldberg, 2021). Of note, with SATs, one can only look at immediate antitumor activities such as responses and cannot make any inferences regarding stable disease which may reflect the natural history of the patients. Endpoints such as disease stabilization or time to event (progression or survival) can only be demonstrated in RCTs.

## 5.3 PI3K inhibitor for hematologic malignancies

Phosphoinositide 3-kinases (PI3Ks) are a class of enzymes regulating multiple cellular processes including cell growth, proliferation, differentiation, survival and intracellular trafficking. Activation of the PI3K signaling pathway is often detected in hematologic malignancies and dysregulated PI3K signaling promotes the survival and proliferation of malignant lymphocytes, which provides the rationale for therapeutic targeting of PI3K isoforms in haematologic malignancies (Richardson et al., 2022).

Given the above background and the unmet medical needs in some rare types of (relapsed and refractory) hematologic malignancies (D'Sa and Zucca, 2024), the FDA granted the first AA in July 2014 of idelalisib to treat patients with relapsed follicular lymphoma (FL), small lymphocytic lymphoma (SLL), and chronic lymphocytic leukemia (CLL). Since then, the agency approved three additional PI3K inhibitors for hematologic cancer indications. The development history and relevant information for each of the approved PI3K inhibitors are summarized as follows (FDA, 2022).

*Idelalisib.* The FDA granted AA in July 2014 to idelalisib, a PI3K$\delta$ inhibitor, in relapsed FL and SLL after ≥ 2 systemic therapies based on an SAT (Study 101-09, NCT01282424) with an ORR of 54% (95% CI 42%–66%) in FL and 58% (95% CI 37%–77%) in SLL. At the same time, the agency also approved idelalisib in combination with rituximab in relapsed

CLL based on an RCT (Study GS-US-312-0116, NCT01539512) with a PFS HR 0.18 (95% CI 0.10–0.31). Three subsequent RCTs in CLL or indolent non-Hodgkin lymphoma (iNHL) were halted in 2016 due to increased deaths and serious toxicities, from which a pooled analysis showed a death rate 7.4% in idelalisib arms versus 3.5% in control arms (OS HR 2.29 with 95% CI: 1.26–4.18). In January 2022, the sponsor voluntarily withdrew FL and SLL, citing poor enrollment in postmarketing requirement (PMR) studies after initial AA (Banerjee et al., 2023)

*Duvelisib.* Duvelisib, a PI3K$\delta$ and PI3K$\gamma$ inhibitor, was granted AA in R/R FL after ≥ 2 systemic therapies based on an SAT (NCT02204982) with an ORR 42% (95% CI 31%–54%) and regular approval in R/R CLL and SLL after ≥ 2 therapies based on a phase III DUO RCT (NCT02004522) comparing duvelisib versus ofatumumab with PFS HR 0.52 (95% CI 0.39–0.69). The final 5-year data of DUO concluded an OS, a key secondary endpoint, of 1.09 (95% CI 0.69–1.51), favoring the control arm. Of note, the DUO study also enrolled patients with a prior line therapy, in which death rates were higher with duvelisib than ofatumumab in both patient groups who received 1 prior therapy (50% versus 44%) and 2 or more prior therapies ((55.8% vs 48.5%). The sponsor voluntarily withdrew R/R FL in December 2021 with the considerations of treatment landscape change after duvelisib gaining AA 4 years ago and timing of a confirmatory study.

*Copanlisib.* The FDA granted AA in September 2017 to copanlisib, a PI3K$\alpha$ and PI3K$\delta$ inhibitor, in relapsed FL based on an SAT (CHRONOS-1, NCT01660451) with ORR 59% (95% CI 49%–68%). A subsequent PMR RCT (CHRONOS-3, (NCT02367040) comparing copanlisib+rituximab versus placebo+rituximab resulted in a PFS HR 0.52 (95% CI 0.39–0.69) and a OS HR 0.87 (95% CI 0.57–1.35) in iNHL and 0.95 (95% CI 0.52–1.74) in FL. In addition, the RCT also found a higher death rate due to SAEs in the copanlisib arm. Consequently, the sponsor voluntarily withdrew in December 2021 the NDA based on CHRONOS-3.

*Umbralisib.* The FDA granted AA in February 2021 to umbralisib, a dual inhibitor of PI3K$\delta$ and casein kinase-1$\epsilon$ (CK1$\epsilon$), in R/R FL after ≥ 3 systemic therapies and R/R marginal zone lymphoma (MZL) after ≥ 1 anti-CD20-based therapy based on an SAT (NCT02793583) with an ORR 43% (95% CI 34%–52%) in FL and 49% (95% CI 37%–62%) in MZL. Data from a phase III RCT (NCT02612311) comparing umbralisib+ublituximab versus obinutuzumab+chlorambucil in previously treated and R/R CLL showed an HR of 0.55 (95% CI 0.41–0.72) in PFS and of 1.10 (95% CI 0.75–1.59) in OS (interim). Patients receiving the novel combination experienced more SAEs, leading to a higher incidence of deaths in umbralisib arm, which prompted the FDA to place multiple clinical trials involving ublituximab and umbralisib on hold. In April 2022, the sponsor voluntarily withdraw umbralisib for the indications FL and MZL.

In summary, the main concerns in the above presented PI3K inhibitors as a class of drugs are (1) potential detriments in OS for PI3K inhibitor arm due to increased toxicity, (2) limited dose finding and/or dosing optimization, (3) insufficient exploration of exposure-response relationship in both efficacy and safety, (4) no or limited dose finding conducted for use in combination, and (5) high rates of treatment modifications due to toxicity.

# 6 Concluding Remarks

There is a growing tendency in oncology drug development and approval pathway moving from SATs towards RCTs. Nevertheless, there are occasions where SATs are appropriate to support regulatory decision. This paper is intended to describe (1) the necessary conditions under which an SAT may be more appropriate, (2) desirable considerations that can make the design of the SAT more optimal or its results more interpretable, (3) further considerations for evidence of product effectiveness, and (4) planning/initiation of a confirmatory RCT within reasonable time after AA. In general, an SAT meeting as many of these conditions as possible may have a higher likelihood of getting AA. However, it is strongly recommended that the sponsor discusses with relevant regulatory agencies before initiation of an SAT for the purpose of regulatory decision.

# Acknowledgements

We would like to thank Dr. Zhimin Yang of the Center for Drug Evaluation of the NMPA, the anonymous referees and the Associate Editor for their constructive comments and suggestions that greatly helped improve the presentation of the paper.

28. FDA. Rare Diseases: Natural History Studies for Drug Development. Guidance for Industry (Draft guidance). US Food and Drug Administration, Silver Spring, MD, 2019.

29. Delgado, A and Guddati, AK. Clinical endpoints in oncology–a primer. *American Journal of Cancer Research*, 11(4):1121–1131, 2021.

30. FDA. OCE Rare Cancers Program–Promoting development of new drug and biological products to treat patients with rare cancers. U.S. Department of Health and Human Services, Food and Drug Administration, 2024.

31. NCI. Targeted Therapy to Treat Cancer. U.S. Department of Health and Human Services, National Institutes of Health, National Cancer Institute, 2022.

32. Topalian, SL, Taube, JM, Anders, RA, and Pardoll, DM. Mechanism-driven biomarkers to guide immune checkpoint blockade in cancer therapy. *Nature Reviews Cancer*, 16(5): 275–287, 2016.

33. Sang, W, Shi, M, Yang, J, Cao, J, Xu, L, Yan, D, Song, X, Sun, C, Li, D, and Zhu, F. Combination of Anti-CD19 and Anti-CD20 Chimeric Antigen Receptor T Cells for Relapsed and Refractory Diffuse Larger B Cell Lymphoma: An Open-Label, Single-Arm, Phase I/II Trial. *Blood*, 134:1590, 2019.

34. Fu, Z, Li, S, Han, S, Shi, C, and Zhang, Y. Antibody drug conjugate: the "biological missile" for targeted cancer therapy. *Signal Transduction and Targeted Therapy*, 7(1):93, 2022.

35. Westphalen, CB, Martins-Branco, D, Beal, J, Cardone, C, Coleman, N, Schram, A, Halabi, S, Michiels, S, Yap, C, André, F, et al. The ESMO Tumour-Agnostic Classifier and Screener (ETAC-S): a tool for assessing tumour-agnostic potential of molecularly guided therapies and for steering drug development. *Annals of Oncology*, 35(11):936–953, 2024.

36. FDA. Optimizing the Dosage of Human Prescription Drugs and Biological Products for the Treatment of Oncologic Diseases—Guidance for Industry. U.S. Department of Health and Human Services Food and Drug Administration, 2023b.

37. Yuan, Y, Zhou, H, and Liu, S. Statistical and practical considerations in planning and conduct of dose-optimization trials. *Clinical Trials*, page 17407745231207085, 2024a.